%% file: main.tex
\documentclass{aa}

\usepackage{graphicx} 
\usepackage{txfonts}
\usepackage{hyperref}
\usepackage{gensymb}
\usepackage{bm}
\usepackage{xcolor}
\usepackage{booktabs}

\input{preamble}

\begin{document}

\title{Constraints on primordial non-Gaussianity from \Planck\ PR4 data}

\author{Gabriel Jung\inst{1}
\and Michele Citran\inst{2,3}
\and Bartjan van Tent\inst{2}
\and Léa Dumilly\inst{1}
\and Nabila Aghanim\inst{1}
}
\authorrunning{G. Jung et al.}

\institute{Université Paris-Saclay, CNRS, Institut d’Astrophysique Spatiale, 91405 Orsay, France 
\and
Universit\'e Paris-Saclay, CNRS/IN2P3, IJCLab, 91405 Orsay, France
\and
Universit\'e Paris Cit\'e, CNRS, Astroparticule et Cosmologie, 75013 Paris, France
}

\date{}

\abstract{We perform the first bispectrum analysis of the final \Planck\ release
temperature and E-polarization CMB data, called PR4. We use the binned bispectrum
estimator pipeline that was also used for the previous \Planck\ releases as well as the
integrated bispectrum estimator. We test the standard primordial (local, equilateral and orthogonal) and secondary (lensing, unclustered point sources and CIB) bispectrum shapes.
The final primordial results of the full T+E analysis are 
$\fnl^\mathrm{local} = -0.1 \pm 5.0$, $\fnl^\mathrm{equil} = 6 \pm 46$ and
$\fnl^\mathrm{ortho} = -8 \pm 21$. These results are consistent with previous \Planck\
releases, but have slightly smaller error bars than in PR3, up to $12\%$ smaller for
orthogonal. They represent the best \Planck\ constraints on primordial non-Gaussianity.
The lensing and point source bispectra are also detected, consistent with PR3. We perform
several validation tests and find in particular that the 600 simulations, used to determine
the linear correction term and the error bars, have a systematically low lensing 
bispectrum. We show however that this has no impact on our results.}

\maketitle

\nolinenumbers

\section{Introduction}
\label{sec:introduction}

Primordial non-Gaussianity (PNG) is the field of study of deviations from Gaussianity of the
primordial cosmological fluctuations. A wealth of information is expected to be encoded into
the non-Gaussianity of the fluctuations, in particular regarding the period of the production
of these fluctuations, most generally expected to be a model of inflation \citep[see e.g.~the reviews][]{Bartolo:2004if,Chen:2010xka,Wang:2013zva,Renaux-Petel:2015bja,Achucarro:2022qrl}. For example, the
size (parametrized by the so-called $\fnl$ parameters) and type of non-Gaussianity, once 
detected, will allow us to distinguish between inflation
models with multiple fields \citep[see e.g.][]{Vernizzi:2006ve,Rigopoulos:2005us,Byrnes:2010em,Jung:2016kfd,dePutter:2016trg,Wang:2022eop}, models with non-standard kinetic terms \citep[see e.g.][]{Seery:2005wm,Chen:2006nt,Bartolo:2010bj}, or the standard single-field
slow-roll models of inflation~\citep{Maldacena:2002vr}. PNG can also put constraints on other models of the early universe 
than inflation, see e.g.~\cite{Lehners:2010fy,Agullo:2017eyh,vanTent:2022vgy}. Together with the tensor-to-scalar ratio, PNG 
is  the main primordial observable that we hope to measure in the future. The tensor-to-scalar 
ratio will tell us the energy scale at which inflation happened and, together with the already 
observed amplitude and spectral index (slope) of the scalar power spectrum, allow us 
to determine also the first and second derivatives of the inflaton potential at that scale, at 
least for single-field slow-roll models~\citep[see e.g.][for a review of inflation theory]{Baumann:2009ds}. However, as indicated above, non-Gaussianity contains much more information for 
distinguishing between inflation models and, contrary to the tensor-to-scalar ratio which could 
be arbitrarily small and hence unobservable, it has an explicit lower limit that is the 
prediction of single-field slow-roll inflation \citep{Maldacena:2002vr} (although that limit is about two orders of
magnitude smaller than what we can hope to measure with the CMB alone). In addition, PNG could
show the presence of new massive particles in the context of the so-called cosmological collider 
\citep{Arkani-Hamed:2015bza}.

It is hence no wonder that PNG was an important focus of cosmological surveys, mainly the cosmic microwave background (CMB) satellite missions of the past, {\em WMAP} \citep[see e.g.][]{WMAP:2010qai,WMAP:2012fli} and \Planck, 
and remains so for future missions like {\em LiteBIRD} \citep{LiteBIRD:2022cnt} 
\citep[see also e.g.][]{CORE:2016ymi}.
In particular, the \Planck\ collaboration dedicated a
full paper to PNG at each of its three official releases (PR1, PR2, and PR3),
\cite{Planck:2013wtn, Planck:2015zfm, Planck:2019kim}. The third of these has the most
stringent observational constraints to date on a large number of types of non-Gaussianity.
A similar level of constraints should in principle be reached or even improved upon with the new generation of galaxy surveys like \textit{Euclid} \citep{EUCLID:2011zbd}, see e.g.~\cite{Karagiannis:2018jdt}, but currently they are not yet competitive with the CMB results \citep[see][for constraints based on the \textit{BOSS} data]{DAmico:2022gki, Cabass:2022wjy,Cabass:2022ymb, Ivanov:2024hgq, Cagliari:2025rqe}.
To determine the CMB bispectrum (the Fourier or spherical harmonic transform of the 
three-point correlator), which is the object where one generically expects the strongest PNG 
signal and the exclusive focus of this paper, three main estimators were used in the \Planck\
papers: the KSW estimator \citep{Komatsu:2003iq,Yadav:2007rk,Yadav:2007ny}, the binned estimator 
\citep{Bucher:2009nm, Bucher:2015ura}, and the modal estimator 
\citep{Fergusson:2009nv,Fergusson:2010dm,Fergusson:2014gea}. 

After the third \Planck\ release, a part of the collaboration continued to work on 
the data-processing pipeline, creating a new one called NPIPE, which led to a final 
release, PR4 \citep{Planck:2020olo}. However, this
release of data products was not accompanied by the usual suite of analysis papers. The
cosmological parameters determined from the power spectra of PR4 were finally published in
\cite{Tristram:2023haj}. However, no non-Gaussian analysis had been performed on the PR4 
data so far.\footnote{While finalizing this work, a paper appeared
on the arXiv using the PR4 data to study the trispectrum (four-point correlator): \cite{Philcox:2025wts}.} In addition to including previously neglected data from the repointing periods, 
NPIPE processed all the \Planck\ channels within the same framework, including the 
latest versions of corrections for systematics and data treatment. This means lower 
noise levels and fewer systematics in PR4 compared to PR3.

In this work we perform the first bispectrum analysis of the PR4 data. We use both the 
effectively optimal (meaning optimal given the error bars) binned bispectrum estimator 
that was also used for all previous \Planck\ releases, and an additional estimator that was 
developed later: the integrated bispectrum estimator \citep{Jung:2020zne}. The integrated 
bispectrum estimator is non-optimal and has only been developed for temperature, but it is 
very fast and allows us to cross-check results and do additional tests. We analyze the
standard primordial (local, equilateral and orthogonal) and secondary (lensing, 
unclustered point sources and CIB) bispectrum shapes. We perform various validation tests
and in particular look very carefully at the 600 CMB and noise simulations that are used
both for the linear correction term and for the error bars.

The outline of the paper is as follows. In section~\ref{sec:bispectrum} we define the CMB 
bispectrum and in particular its binned and integrated approximations. In section~\ref{sec:parest} we recall the theoretical templates we will test for
and the expression of the $\fnl$ estimators. In section~\ref{sec:data} we describe 
the data and simulations we use. The main results of our paper are given in section~\ref{sec:results}, 
followed by section~\ref{sec:tests} with validation tests to confirm their robustness. 
We conclude in section~\ref{sec:conclusions}.

\section{The CMB bispectrum}
\label{sec:bispectrum}

\subsection{Definitions}

To study small deviations from Gaussianity in the CMB, a key observable is the angle-averaged bispectrum defined by
\begin{equation}
  \label{eq:bispectrum-definition}
  B_{\ell_1 \ell_2 \ell_3}^{p_1 p_2 p_3} \equiv 
  \left\langle \int_{S^2} \mathop{}\!\mathrm{d} \hat\Omega \, M_{\ell_1}^{p_1}(\hat\Omega) M_{\ell_2}^{p_2}(\hat\Omega) M_{\ell_3}^{p_3}(\hat\Omega)\right\rangle,
\end{equation}
where maps of the CMB temperature or E-polarization field denoted by the label $p$ are filtered using the standard decomposition into spherical harmonics:
\begin{equation}
    \label{eq:filtered-maps}
    M_\ell^p(\hat{\Omega}) = 
    \sum\limits_{m=-\ell}^{+\ell} a_{\ell m}^{p} Y_{\ell m}(\hat{\Omega}).
\end{equation}

One can show that this angle-averaged bispectrum is related to the full angular bispectrum, which is defined as the three-point correlator of the harmonic coefficients, under the assumption of statistical isotropy by
\begin{equation}
    \label{eq:angular-bispectrum}
      B_{\ell_1 \ell_2 \ell_3} = h_{\ell_1 \ell_2 \ell_3} \sum\limits_{m_1, m_2, m_3}
  \begin{pmatrix}
    \ell_1 & \ell_2 & \ell_3\\
    m_1 & m_2 & m_3
  \end{pmatrix}
  \langle a_{\ell_1 m_1} a_{\ell_2 m_2} a_{\ell_3 m_3}\rangle,
\end{equation}
where 
\begin{equation}
  h_{\ell_1 \ell_2 \ell_3} = 
  \sqrt{\frac{(2  \ell_1+1) (2  \ell_2+1) (2  \ell_3+1)}{4\pi}}  
  \begin{pmatrix}
     \ell_1 &  \ell_2 &  \ell_3\\
    0 & 0 & 0
  \end{pmatrix}.
\end{equation}
Due to the presence of the $3j$ Wigner symbol in $h_{\ell_1 \ell_2 \ell_3}$, we need to consider only $\ell$-triplets that respect both the parity condition ($\ell_1+\ell_2+\ell_3$ even) and the triangle inequality ($|\ell_1-\ell_2|\leq\ell_3\leq\ell_1+\ell_2$).

Assuming weak non-Gaussianity, the bispectrum covariance matrix in polarization space becomes
\begin{equation}
  \label{eq:bispectrum-variance}
  \mathrm{Covar}(B_{\ell_1 \ell_2 \ell_3}^{p_1 p_2 p_3}, B_{\ell_1 \ell_2 \ell_3}^{p_4 p_5 p_6}) = 
  g_{\ell_1 \ell_2 \ell_3} h_{\ell_1 \ell_2 \ell_3}^2 C_{\ell_1}^{p_1 p_4} C_{\ell_2}^{p_2 p_5} C_{\ell_3}^{p_3 p_6},
\end{equation}
where the factor $g_{\ell_1 \ell_2 \ell_3}$ is $6$ when the three $\ell$'s are the same, $2$ when only two are equal and $1$ otherwise.

In practice, measuring the bispectrum for all valid configurations is computationally prohibitive. To address this challenge, several approaches have been developed, including the binned bispectrum and integrated bispectrum estimators, which we discuss in the following.

\subsection{Binned bispectrum}
\label{sec:binned-bispectrum}

The binned bispectrum method simply consists of using a binning of the multipole $\ell$ space to drastically reduce the number of mode triplets. It is well adapted to work with relatively smooth functions in harmonic space, which is the case of the different standard bispectrum templates we discuss later in Sect.~\ref{sec:shapes}. The binned bispectrum estimator and its implementation are described in \cite{Bucher:2015ura} \citep[see also][for details]{Bucher:2009nm, Jung:2018rgf}.

The binned bispectrum estimator can be written as
\begin{equation}
  \label{eq:binned-bispectrum-estimator}
  \hat{B}_{i_1 i_2 i_3}^{p_1 p_2 p_3} =
  \int_{S^2} \mathop{}\!\mathrm{d} \hat\Omega \, M_{i_1}^{p_1}(\hat\Omega) M_{i_2}^{p_2}(\hat\Omega) M_{i_3}^{p_3}(\hat\Omega) 
  - B_{i_1 i_2 i_3}^{p_1 p_2 p_3, \mathrm{lin}},
\end{equation}
where maps filtered over the $i$-th bin, denoted as $\Delta_i$, are simply obtained using
\begin{equation}
\label{eq:binned-filtered-maps}
    M_i^p(\hat{\Omega}) = \sum\limits_{\ell\in\Delta_i}M_\ell^p(\hat{\Omega}).
\end{equation}
The linear correction term $B_{i_1 i_2 i_3}^{p_1 p_2 p_3, \mathrm{lin}}$ is necessary to maintain the optimality of the estimator when features breaking the expected isotropy of the signal have to be taken into account. For example with observational datasets, masking some regions of the sky like the galactic plane as well as the anisotropic scanning pattern of the mission which leads to anisotropic noise levels, have such an effect. This correction is computed using
\begin{equation}
    \label{eq:binned-bispectrum-linear-correction}
    \begin{split}
    B_{i_1 i_2 i_3}^{p_1 p_2 p_3, \mathrm{lin}} = \int_{S^2} \mathop{}\!\mathrm{d} \hat\Omega \, 
    \Big[&M_{i_1}^{p_1}\left\langle M_{i_2}^{p_2} M_{i_3}^{p_3}\right\rangle \\
     &+ M_{i_2}^{p_2}\left\langle M_{i_1}^{p_1} M_{i_3}^{p_3}\right\rangle + M_{i_3}^{p_3}\left\langle M_{i_1}^{p_1} M_{i_2}^{p_2}\right\rangle \Big],
    \end{split}
\end{equation}
where the ensemble averages are calculated over CMB simulations with the same characteristics as the observations (beam, noise and mask). Moreover, the estimated binned bispectrum should be multiplied by the factor $1/\fsky$, with $\fsky$ the unmasked fraction of the sky, before any comparison to full-sky theoretical predictions.

From a known theoretical bispectrum template (see Sect.~\ref{sec:shapes} for examples), it is straightforward to compute the corresponding theoretical binned bispectrum
\begin{equation}
    \label{eq:binned-bispectrum-theory}
    B_{i_1 i_2 i_3}^{p_1 p_2 p_3,\mathrm{th}} = \sum\limits_{\ell_1\in\Delta_1}\sum\limits_{\ell_2\in\Delta_2}\sum\limits_{\ell_3\in\Delta_3} B_{\ell_1 \ell_2 \ell_3}^{p_1 p_2 p_3, \mathrm{th}}.
\end{equation}
The covariance matrix of the binned bispectrum in polarization space is given by
\begin{equation}
  \label{eq:binned-bispectrum-variance}
  \begin{split}
  \mathrm{Covar}(B_{i_1 i_2 i_3}^{p_1 p_2 p_3}, B_{i_1 i_2 i_3}^{p_4 p_5 p_6}) &\equiv V^{p_1 p_2 p_3 p_4 p_5 p_6}_{i_1 i_2 i_3}\\
  &= 
  g_{i_1 i_2 i_3} \\
&\times\sum\limits_{\ell_1\in\Delta_1}\sum\limits_{\ell_2\in\Delta_2}\sum\limits_{\ell_3\in\Delta_3}
  h_{\ell_1 \ell_2 \ell_3}^2 C_{\ell_1}^{p_1 p_4} C_{\ell_2}^{p_2 p_5} C_{\ell_3}^{p_3 p_6}.
  \end{split}
\end{equation}
To take into account instrumental effects, one has to apply the following modifications to Eqs.~\eqref{eq:binned-bispectrum-theory} and \eqref{eq:binned-bispectrum-variance}
\begin{equation}
    \label{eq:instrumental-effects}
    C_\ell \rightarrow b_\ell^2 C_\ell + N_\ell, \qquad 
    B_{\ell_1 \ell_2 \ell_3} \rightarrow b_{\ell_1} b_{\ell_2} b_{\ell_3} B_{\ell_1 \ell_2 \ell_3},
\end{equation}
where $b_\ell$ is the beam window function and $N_\ell$ the noise power spectrum. 

\subsection{Integrated bispectrum}
\label{sec:integrated-bispectrum}

The integrated bispectrum method follows a different and simpler approach to extract the relevant bispectral information. Based on the original idea of \cite{Chiang:2014oga}, developed in the context of large-scale structure study, it consists of dividing the sky into many equal-size patches and computing the average correlation of the mean value and the power spectrum in each patch. We adopt the same methodology as in \cite{Jung:2020zne}, recalling here only the key steps.

The full integrated bispectrum estimator can be written as
\begin{equation}
    \label{eq:integrated-bispectrum-estimator}
    \hat{\ib}_\ell = \frac{1}{N_\mathrm{patch}}\sum\limits_{\mathrm{patch}}
        \overline{M}^\mathrm{patch}C_{\ell}^\mathrm{patch}\, - \ib_\ell^\mathrm{lin},
\end{equation}
where $\overline{M}^\mathrm{patch}$ and $C_{\ell}^\mathrm{patch}$ are respectively the measured mean value and power spectrum in a given patch, and $N_\mathrm{patch}$ is the total number of patches. Note that in this expression, we drop the $p$ subscript, as we only consider temperature maps with the integrated bispectrum estimator in this work. Similarly to the binned bispectrum case, the linear correction term $\ib_\ell^\mathrm{lin}$ is used to maintain optimality when anisotropy breaking effects (mask, anisotropic noise) are present in the data. It takes the form
\begin{equation}
    \label{eq:integrated-bispectrum-linear-correction}
    \ib_\ell^\mathrm{lin} = \frac{1}{N_\mathrm{patch}}\sum\limits_{\mathrm{patch}}
        \overline{M}^\mathrm{patch}C_{\ell}^\mathrm{patch,sims},
\end{equation}
where the $C_{\ell}^\mathrm{patch,sims}$ are evaluated from a sufficiently large set of simulations reproducing the main characteristics of observations (e.g., mask, noise and beam). Additionally, the estimated integrated bispectrum is multiplied by the factor $1/\fsky$ if the sky is partially masked.

One can directly relate the theoretical expectation of the integrated bispectrum and the theoretical bispectrum template, and the expression only depends on the shape, size, and position of the patches. In the case of azimuthally symmetric patches, this link is given by:
\begin{equation}
    \label{eq:integrated-bispectrum-theory}
        \ib_\ell^\mathrm{th} 
        = \frac{1}{(4\pi)^3} \sum\limits_{\ell_1 \ell_2 \ell_3 \ell_4 \ell_5} w_{\ell_3} w_{\ell_4} w_{\ell_5} \mathcal{F}_{\ell\ell_1 \ell_2 \ell_3 \ell_4 \ell_5} B_{\ell_1 \ell_2 \ell_3}^\mathrm{th}  \,, \\
\end{equation}
where the factor
\begin{equation}
    \begin{split}        
    \mathcal{F}_{\ell\ell_1 \ell_2 \ell_3 \ell_4 \ell_5} &\equiv (-1)^{\ell_2 + \ell_4} (2\ell_4+1)(2\ell_5+1) \begin{pmatrix}
            \ell_1 &  \ell_2 &  \ell_3\\
            0  & 0 & 0
        \end{pmatrix}^{-1} \\
        &\times
        \begin{pmatrix}
            \ell &  \ell_1 &  \ell_4\\
            0  & 0 & 0
        \end{pmatrix}
        \begin{pmatrix}
            \ell &  \ell_2 &  \ell_5\\
            0  & 0 & 0
        \end{pmatrix}
        \begin{pmatrix}
            \ell_3 &  \ell_4 &  \ell_5\\
            0  & 0 & 0
        \end{pmatrix}
        \begin{Bmatrix}
            \ell_1 &  \ell_2 &  \ell_3\\
            \ell_5  & \ell_4 & \ell
        \end{Bmatrix}
    \end{split}
\end{equation}
only depends on Wigner $3j$ and $6j$ symbols (in parentheses and braces, respectively), and $w_\ell$ is a simple function in harmonic space that defines the full set of patches.
\footnote{The real-space map of a chosen azimuthally symmetric patch centered at $\hat{\Omega}^\mathrm{patch}$ is given by $W(\hat{\Omega},\hat{\Omega}^\mathrm{patch})=\sum\limits_{\ell} w_\ell \frac{2\ell + 1}{4\pi} P_\ell(\hat{\Omega}\cdot\hat{\Omega}^\mathrm{patch})$ where $P_\ell$ is a Legendre polynomial.} 
In this work, we use the same set of patches as in \cite{Jung:2020zne} defined by $w_\ell=1$ for $\ell \leq 10$ and $0$ otherwise. This implies that we probe only the squeezed limit of the bispectrum.

The covariance of the integrated bispectrum can be computed in a similar manner using the weakly non-Gaussian approximation and Eq.~\eqref{eq:bispectrum-variance}:
\begin{equation}
    \label{eq:integrated-bispectrum-covariance}
    \begin{split}
        \ic_{\ell\ell'}  \equiv & \, \mathrm{Covar}(\ib_{\ell},\ib_{\ell'})\\
        = & \frac{1}{(4\pi)^6} \sum\limits_{\ell_1 \ell_2 \ell_3 \ell_4 \ell_5  \ell'_4 \ell'_5}   h_{\ell_1 \ell_2 \ell_3}^2 \mathcal{F}_{\ell\ell_1 \ell_2 \ell_3 \ell_4 \ell_5} w_{\ell_3} w_{\ell_4} w_{\ell_5} w_{\ell'_4} w_{\ell'_5} C_{\ell_1}C_{\ell_2}C_{\ell_3} \\
        & \times \left[w_{\ell_3} (\mathcal{F}_{\ell'\ell_1 \ell_2 \ell_3 \ell'_4 \ell'_5} + \mathcal{F}_{\ell'\ell_2 \ell_1 \ell_3 \ell'_4 \ell'_5}) \right.\\
        & \, + w_{\ell_2} (\mathcal{F}_{\ell'\ell_1 \ell_3 \ell_2 \ell'_4 \ell'_5} + \mathcal{F}_{\ell'\ell_3 \ell_1 \ell_2 \ell'_4 \ell'_5}) \\
        & \, \left. + w_{\ell_1} (\mathcal{F}_{\ell'\ell_3 \ell_2 \ell_1 \ell'_4 \ell'_5} + \mathcal{F}_{\ell'\ell_2 \ell_3 \ell_1 \ell'_4 \ell'_5})\right].
    \end{split}
\end{equation}
In the realistic case, the same modifications as for the binned bispectrum (Eq.~\ref{eq:instrumental-effects}) should be applied to Eqs.~\eqref{eq:integrated-bispectrum-theory} and \eqref{eq:integrated-bispectrum-covariance}.

\section{Parametric estimation}
\label{sec:parest}

\subsection{Theoretical shapes}
\label{sec:shapes}

The presence of a small amount of PNG is a prediction of many different inflationary models beyond the standard single-field slow-roll scenario. These models produce distinct bispectrum shapes depending on their specific characteristics. Here, we focus on the local, equilateral, and orthogonal templates that were studied in detail in the \Planck\ PNG analyses \citep{Planck:2013wtn, Planck:2015zfm, Planck:2019kim}.

These three bispectrum shapes can be written as functions of the primordial power spectrum of the gravitational potential $P(k)=A(k/k_0)^{n_\mathrm{s}-4}$, where $A$ is its amplitude, $k_0$ the pivot scale and $n_\mathrm{s}$ the spectral index,
\begin{equation}
    \label{eq:bispectrum-local-equil-ortho}
    \begin{split}
      B^\mathrm{local}(k_1,k_2,k_3) &= 2 [ P(k_1) P(k_2) + P(k_1) P(k_3) + P(k_2) P(k_3) ],\\
      B^\mathrm{equil}(k_1,k_2,k_3) &= - 6 [ P(k_1)P(k_2) + (\mathrm{2\ perms})] \\
      &\quad - 12 \,P^{2/3}(k_1) P^{2/3}(k_2) P^{2/3}(k_3) \\
      &\quad +6 [ P(k_1) P^{2/3}(k_2) P^{1/3}(k_3) + (\mathrm{5\ perms})],\\
      B^\mathrm{ortho}(k_1,k_2,k_3) &=  -18 [ P(k_1)P(k_2) + (\mathrm{2\ perms}) ] \\
      &\quad - 48 \,P^{2/3}(k_1) P^{2/3}(k_2) P^{2/3}(k_3)\\
      &\quad +18 [ P(k_1) P^{2/3}(k_2) P^{1/3}(k_3) + (\mathrm{5\ perms}) ].
    \end{split}
\end{equation}
The local type of PNG \citep{Gangui:1993tt} is mostly related to multiple-field inflation, where curvature perturbations can evolve on superhorizon scales \citep[see e.g.][]{Rigopoulos:2005us,Byrnes:2010em}. The equilateral and orthogonal templates cover a wide range of inflationary models that have non-standard kinetic terms or higher-derivative interactions \citep[see e.g.][]{Creminelli:2005hu, Chen:2006nt, Senatore:2009gt}.

The corresponding bispectra in the CMB anisotropies can then be computed using \citep{Komatsu:2001rj}
\begin{equation}
    \begin{split}
  \label{eq:bispectrum-primordial}
B_{\ell_1 \ell_2 \ell_3}^{p_1 p_2 p_3, \mathrm{th}} =
&h^2_{\ell_1 \ell_2 \ell_3}
\left( \frac{2}{\pi} \right)^3
\int _0^\infty \!\!\! k_1^2 dk_1 \int _0^\infty \!\!\! k_2^2 dk_2
\int _0^\infty \!\!\! k_3^2 dk_3 \\
& 
\Bigl[
\Delta_{\ell _1}^{p_1}(k_1) \Delta_{\ell _2}^{p_2}(k_2) \Delta_{\ell_3}^{p_3}(k_3)
B(k_1,k_2,k_3)
\\
& 
\times \int _0^\infty \!\!\! r^2dr \, j_{\ell _1}(k_1r) j_{\ell _2}(k_2r)
j_{\ell _3}(k_3r) \Bigr],
\end{split}
\end{equation}
where $\Delta_{\ell}^{p}(k)$ are the radiation transfer functions and $j_\ell$ are the spherical Bessel functions.

There are other sources of bispectral NG in the CMB due to the presence of different foregrounds. The contribution of the three main ones, that are observable in the \Planck\ data, can be taken into account using the following templates.

First, the bispectrum due to gravitational lensing (in temperature caused by the correlation of the integrated Sachs-Wolfe (ISW) and lensing effects) \citep{Hu:2000ee, Lewis:2011fk} is
\begin{equation}
  \label{eq:bispectrum-ISW-lensing}
B_{\ell_1 \ell_2 \ell_3}^{p_1 p_2 p_3, \mathrm{lens}} = h_{\ell_1 \ell_2 \ell_3}^2 \Bigl[
f_{\ell_1 \ell_2 \ell_3}^{p_1}C_{\ell_2}^{p_2\phi}C_{\ell_3}^{p_1 p_3} +  (\mathrm{5\ perms})\Bigr],
\end{equation}
where the $C_{\ell}^{p_1 p_2}$'s are lensed power spectra and $C_{\ell}^{T\phi}$ ($C_{\ell}^{E\phi}$) is the temperature(polarization)-lensing cross-power spectrum. The $f_{\ell_1 \ell_2 \ell_3}^{p}$ factors are given by
\begin{equation}
\begin{split}
f_{\ell_1 \ell_2 \ell_3}^{T} = &\frac{1}{2}\left[ \ell_2(\ell_2 + 1) + \ell_3(\ell_3 + 1) - \ell_1(\ell_1 + 1)\right],\\
f_{\ell_1 \ell_2 \ell_3}^{E} = &\frac{1}{2}\left[ \ell_2(\ell_2 + 1) + \ell_3(\ell_3 + 1) - \ell_1(\ell_1 + 1)\right]\\
&\times
\begin{pmatrix}
 \ell_1 &  \ell_2 &  \ell_3 \\
 2 & 0 & -2
\end{pmatrix}
\begin{pmatrix}
 \ell_1 &  \ell_2 &  \ell_3\\
0 & 0 & 0
\end{pmatrix}^{-1}
.
\end{split}
\end{equation}

The unclustered point sources NG can be described by the simple shape \citep{Komatsu:2001rj}
\begin{equation}
    \label{eq:bispectrum-unclustered}
    B_{\ell_1 \ell_2 \ell_3}^\mathrm{ps} = h_{\ell_1 \ell_2 \ell_3}^2,
\end{equation}
while the cosmic infrared background (CIB) contribution is given by the heuristic template \citep{Lacasa:2013yya, Penin:2013zya}
\begin{equation}
    \label{eq:CIB-shape}
    B_{\ell_1 \ell_2 \ell_3}^\mathrm{CIB} =
    h_{\ell_1 \ell_2 \ell_3}^2
    \left[ \frac{(1+\ell_1/\ell_\mathrm{break}) (1+\ell_2/\ell_\mathrm{break})
    (1+\ell_3/\ell_\mathrm{break})}{(1+\ell_0/\ell_\mathrm{break})^3}\right]^q,
\end{equation}
where the break and pivot scales are $\ell_\mathrm{break}=70$ and $\ell_0=320$, respectively, and the index is $q=0.85$.

\subsection{Estimator}

The optimal estimator of the amplitude parameter $\fnl$ for a given bispectrum shape $B^\mathrm{th}$ determined from the observed
bispectrum $\hat{B}$ is
\begin{equation}
    \label{eq:fnl-estimator}
    \hat{f}_\mathrm{NL} = \frac{\langle \hat{B},B^\mathrm{th} \rangle}{\langle B^\mathrm{th},B^\mathrm{th} \rangle},
\end{equation}
where the exact definition of the inner product denoted as $\langle x,y \rangle$ depends on the bispectrum estimator used. For the binned and integrated bispectra, respectively, it is
\begin{equation}
    \label{eq:inner-products}
    \begin{split}
    &\langle B^a, B^b \rangle^\mathrm{binned} \equiv \sum\limits_{i_1\leq i_2\leq i_3} \sum\limits_{\substack{p_1 p_2 p_3\\   p_4 p_5 p_6}} B_{i_1 i_2 i_3}^{p_1 p_2 p_3,\,a} \left(V^{-1}\right)^{p_1 p_2 p_3 p_4 p_5 p_6}_{i_1 i_2 i_3} B_{i_1 i_2 i_3}^{p_4 p_5 p_6,\,b}, \\
    &\langle B^a, B^b \rangle^\mathrm{integrated} \equiv \sum\limits_{\ell_1\ell_2} 
    \ib_{\ell_1}^{a}\left(\ic^{-1}\right)_{\ell_1\ell_2}\ib_{\ell_2}^{b}.
    \end{split}
\end{equation}
The expected variance of $\hat{f}_\mathrm{NL}$ is given by $1/\langle B^\mathrm{th},B^\mathrm{th} \rangle$, which has to be multiplied by $1/\fsky$ in the case of incomplete sky coverage.

Another important quantity is the bias $\fnl^{(1),(2)}$ on the measurement of $\fnl^{(1)}$ for a given shape $B^{(1)}$ due to the presence of another bispectrum $B^{(2)}$ of known amplitude $\fnl^{(2)}$:
\begin{equation}
    \label{eq:fnl-bias}
    \fnl^{(1),(2)} = - \frac{F_{12}}{F_{11}} \fnl^{(2)},
\end{equation}
where the Fisher matrix elements are $F_{ab}=\langle B^a, B^b \rangle$.
This is relevant to subtract the impact of the fully known lensing bispectrum (see Eq.~\eqref{eq:bispectrum-ISW-lensing}; its
$\fnl$ is known to be unity) on the PNG estimation.

If the amplitude of the other shape(s) is not known, the correct approach is to do a joint estimation of the different shapes of interest. In that case, the estimator defined in Eq.~\eqref{eq:fnl-estimator} becomes
\begin{equation}
    \label{eq:fnl-estimator-joint}
    \hat{f}_\mathrm{NL}^{a} = \sum\limits_{b} (F^{-1})_{ab} \langle \hat{B},B^{b} \rangle.
\end{equation}
In this case, the expected variance of $\hat{f}_\mathrm{NL}^{a}$ 
(also called Fisher error bars after taking a square root), which 
in the independent case can be written as $1/F_{aa}$ as indicated 
above, becomes for the joint analysis $(F^{-1})_{aa}$.


\section{Data and simulations}
\label{sec:data}

Our analyses are based on data and simulations from the latest \Planck\ release (PR4) \citep[see][for details]{Planck:2020olo}. We use the temperature and polarization maps produced by the component separation technique \texttt{SEVEM} \citep{Leach:2008fi, Fernandez-Cobos:2011mmt}, which removes the large NG contamination due to several galactic foregrounds \citep[see e.g.][]{Jung:2018rgf, Coulton:2019bnz}. These maps, which have a beam size of 5 arcminutes, are available on the \Planck\ Legacy Archive\footnote{\url{http://pla.esac.esa.int/pla/}} (PLA). We also utilize the $600$ corresponding CMB and noise simulations, which can be downloaded from the National Energy Research Scientific Computing Center\footnote{\url{https://portal.nersc.gov/project/cmb/planck2020/}} (NERSC). 

We use the $2018$ \Planck\ common mask \citep{Planck:2018yye}, also available on the PLA, which leaves an observed fraction of the sky $\fsky=0.779$ and $\fsky=0.781$ for temperature and polarization, respectively. After applying this mask to the different maps, the same standard diffusive inpainting technique as for previous \Planck\ releases \citep{Bucher:2015ura,Gruetjen:2015sta} is used to remove spurious effects due to both the lack of small-scale power inside the mask and the edge discontinuity, both of which would bleed into the unmasked region of the map during the filtering process in harmonic space.

For comparison purposes, we also consider the results obtained previously in \cite{Planck:2019kim} and \cite{Jung:2020zne} on the \texttt{SEVEM} maps (data and $300$ simulations) from the previous release (PR3) \citep[see][for more information]{Planck:2018yye}. The binned bispectrum estimator uses the same binning for the PR4 analysis as was used for PR3, and the integrated bispectrum estimator uses the same set of patches.

\begin{figure*}
    \centering
    \includegraphics[width=0.48\linewidth]{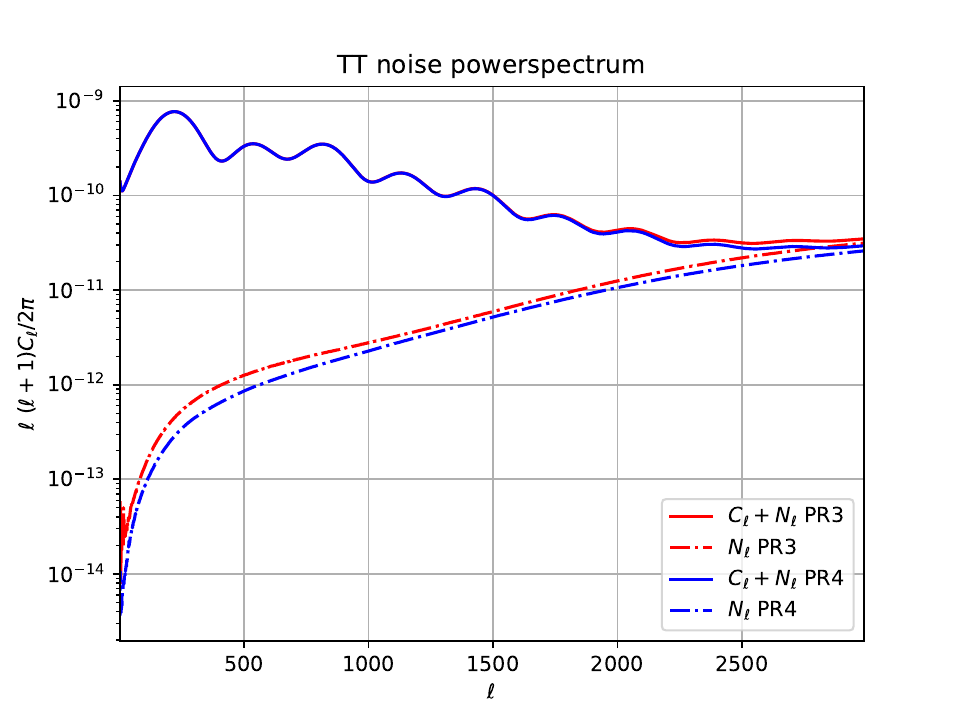}
    \includegraphics[width=0.48\linewidth]{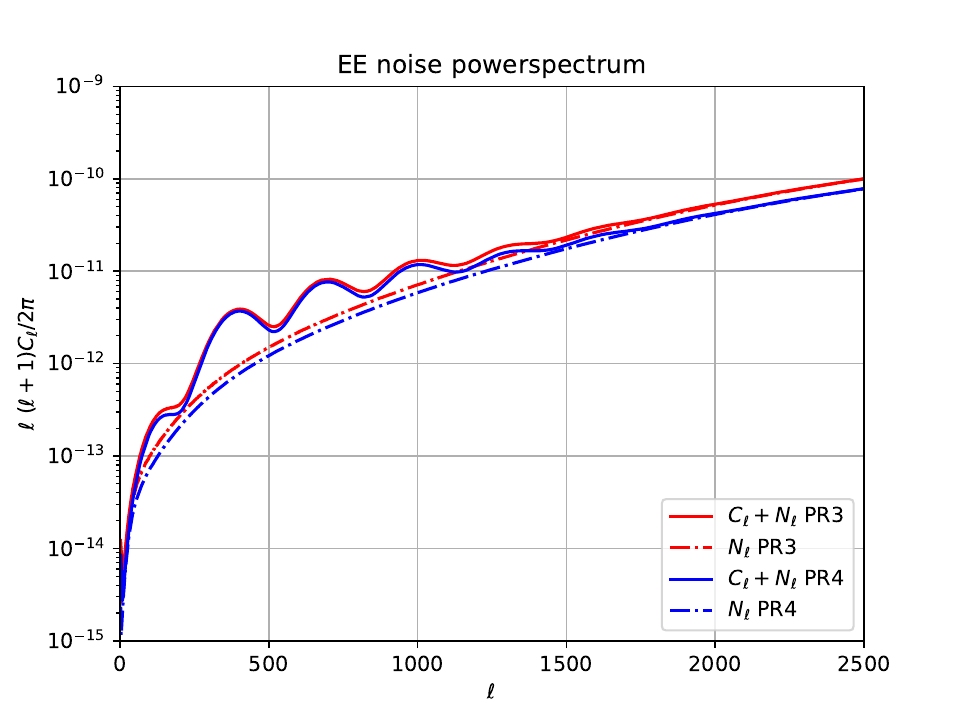}
    \caption{The full and the noise-only power spectra determined from \texttt{SEVEM}
    PR3 and PR4 simulations, for temperature (left) and 
    E-polarization (right).}
    \label{fig:noise}  
\end{figure*}

In Fig.~\ref{fig:noise}, we show the temperature and E-polarization noise power spectra determined from the \texttt{SEVEM} PR3 and PR4 simulations, as well as the full CMB+noise power spectra. This confirms the lower level of noise in the PR4 data on a wide range of scales, allowing for example to extract additional non-Gaussian information from small scales.

\section{Main results}
\label{sec:results}

In this section, we present our analysis of bispectral non-Gaussianity in the latest \Planck\ data. We update the constraints on PNG of the official \Planck\ PNG analyses \citep{Planck:2013wtn, Planck:2015zfm, Planck:2019kim} using a similar methodology. We study the temperature and E-polarization CMB maps produced by the component separation technique \texttt{SEVEM} for the \Planck\ 2020 data release (PR4, see Sect.~\ref{sec:data}). We utilize two different bispectrum estimators to cross-validate our findings, both introduced in Sect.~\ref{sec:bispectrum}. Our main results are obtained by applying to the full dataset the binned bispectrum estimator to investigate standard shapes of primordial origin, namely the local, equilateral and orthogonal ones, and late-time NG in the form of lensing, CIB and unclustered point sources (all described in Sect.~\ref{sec:shapes}). Additionally, we use the integrated bispectrum estimator to constrain the local and lensing bispectra, which both peak in the squeezed limit, from the CMB temperature map alone.

In Table~\ref{tab:png}, we show our constraints on the PNG amplitudes $\fnl^\mathrm{local}$, $\fnl^\mathrm{equil}$ and $\fnl^\mathrm{ortho}$. These results can be compared to Table~5 of \cite{Planck:2019kim} and Table~3 of \cite{Jung:2020zne} for the binned and integrated bispectrum estimators, respectively. These values are obtained in independent estimations, with and without removing the bias due to the expected presence of the lensing bispectrum in the data using Eq.~\eqref{eq:fnl-bias}, and including scales up to $\lmax=2500$. The more conservative approach of analyzing jointly the lensing contribution with the PNG shapes leads to similar conclusions with only slightly larger error bars, as can be seen in Table~\ref{tab:png-joint}. As in the analyses conducted on data from the previous \Planck\ releases, the three considered $\fnl$ parameters are fully compatible with zero in all cases.

\begin{table*}[htbp!]   
\centering
\caption{Constraints on PNG from \Planck\ PR4}
\label{tab:png}
\renewcommand{\arraystretch}{1.2} 
\setlength{\tabcolsep}{8pt} 
\begin{tabular}{l|cccccc}
\toprule\toprule
 & \multicolumn{3}{c}{Independent} & \multicolumn{3}{c}{Lensing bias subtracted} \\
\cmidrule(lr){2-4} \cmidrule(lr){5-7}
Estimator & Local & Equilateral & Orthogonal & Local & Equilateral & 
Orthogonal \\
\midrule
& \multicolumn{6}{c}{$\mathbf{T}$} \\
Binned & $8.9 \pm 5.5$ & $23 \pm 72$ & $-12 \pm 35$ & $1.5 \pm 5.5$ & $23 \pm 72$ & $14 \pm 35$ \\
Integrated & $11.5 \pm 7.3$ & --  & --  & $7.2 \pm 7.3$ & --  & -- \\
\midrule
& \multicolumn{6}{c}{$\mathbf{E}$} \\
Binned & $17 \pm 24$ & $21 \pm 150$ & $-31 \pm 77$ & $16 \pm 24$ & $20 \pm 150$ & $-30 \pm 77$ \\
\midrule
& \multicolumn{6}{c}{$\mathbf{T}$ + $\mathbf{E}$} \\
Binned & $4.9 \pm 5.0$ & $8 \pm 46$ & $-18 \pm 21$ & $-0.1 \pm 5.0$ & $6 \pm 46$ & $-8 \pm 21$ \\
\bottomrule\bottomrule
\end{tabular}
\tablefoot{The constraints on the PNG amplitude parameters $\fnl^\mathrm{local}$, $\fnl^\mathrm{equil}$ and $\fnl^\mathrm{ortho}$ are determined independently (Eq.~\ref{eq:fnl-estimator}) using the binned and integrated bispectrum estimators. Temperature and polarization maps produced by \texttt{SEVEM} from the PR4 dataset are used, including $600$ simulations to evaluate error bars. Constraints are shown both for temperature and polarization independently, and jointly, and without and with taking into account the bias due to the lensing bispectrum.}
\end{table*}

\begin{table*}[htbp!]   
\centering
\caption{Joint constraints on NG from \Planck\ PR4}
\label{tab:png-joint}
\renewcommand{\arraystretch}{1.2} 
\setlength{\tabcolsep}{8pt} 
\begin{tabular}{l|cccccc}
\toprule\toprule
Estimator & Local & Equilateral & Orthogonal & Lensing & Point sources$/10^{29}$ & CIB$/10^{27}$ \\
\midrule
& \multicolumn{6}{c}{$\mathbf{T}$} \\
Binned & $6.7 \pm 6.6 $ & $23 \pm 72$ & $14 \pm 35$ & $0.45 \pm 0.30$ & $7.3 \pm 2.5$  & $0.03 \pm 1.3$ \\
Integrated & $7.2 \pm 8.2$ & --  & --  & $1.01 \pm 1.06$ & --  & -- \\
\midrule
& \multicolumn{6}{c}{$\mathbf{E}$} \\
Binned & $24 \pm 34 $ & $-33 \pm 166 $ & $5 \pm 105 $ & $-4.6 \pm 4.2 $ & -- & -- \\
\midrule
& \multicolumn{6}{c}{$\mathbf{T}$ + $\mathbf{E}$} \\
Binned & $0.5 \pm 6.2 $ & $3 \pm 49 $ & $13 \pm 25 $ & $0.62 \pm 0.23 $ & -- & -- \\
\bottomrule\bottomrule
\end{tabular}
\tablefoot{The constraints on the bispectrum amplitude parameters (both primordial and non-primordial) are determined jointly (Eq.~\ref{eq:fnl-estimator-joint}) using the binned and integrated bispectrum estimators. Temperature and polarization maps produced by \texttt{SEVEM} from the PR4 dataset are used, including $600$ simulations to evaluate error bars. Constraints are shown both for temperature and polarization independently, and jointly.}
\end{table*}

Table~\ref{tab:isw-lensing} summarizes our results for the bispectrum amplitudes from late-time NG, to be compared to Tables~1 and 4 of \cite{Planck:2019kim}. As known from previous analyses, the lensing bispectrum is detected at a value rather low but compatible with the expected amplitude of $1$. As the detected value is extremely close to the \Planck\ $2018$ result, we recall that in \cite{Planck:2019kim} it was pointed out that performing the same analysis on maps where the Sunayev-Zeldovich (SZ) signal has been removed in addition to the galactic foreground contribution increases the measured lensing amplitude to a value much closer to $1$. This is likely due to the correlations between the ISW and tSZ (thermal SZ) effects that produce another bispectrum, as was first shown by \cite{Hill:2018ypf}. These maps have not been produced for the PR4 release, which implies that we could not perform a similar analysis. However, it was verified by \cite{Jung:2020zne} and \cite{Coulton:2022wln} that the ISW-tSZ-tSZ bispectrum amplitude is too small at the \Planck\ resolution to impact significantly our constraints on PNG and we expect the same conclusion to hold here.

The unclustered point sources and CIB amplitudes, $b_\mathrm{PS}$ and $b_\mathrm{CIB}$ respectively, are estimated jointly as these two shapes are highly correlated. The former is detected at a nearly $3\sigma$ level in the data. However, it is important to recall that unlike the lensing bispectrum, this shape has a very low correlation with the various primordial templates, and therefore cannot contaminate the results presented in Table~\ref{tab:png}. The results for these two templates are only shown for temperature. The CIB is not expected to be polarized,
and no polarized version of its template exists. As for unclustered point sources, none are detected in polarization, and as the point sources in temperature and polarization are not necessarily the same, it does not make sense to do a T+E analysis.

\begin{table}[htbp!]
\centering
\caption{Constraints on non-primordial NG from \Planck\ PR4}
\label{tab:isw-lensing}
\renewcommand{\arraystretch}{1.2}
\setlength{\tabcolsep}{8pt} 
\begin{tabular}{l|ccc}
\toprule\toprule
Estimator & lensing & $b_\mathrm{PS}/10^{-29}$ & $b_\mathrm{CIB}/10^{-27}$\\
\midrule
& \multicolumn{3}{c}{$\mathbf{T}$} \\
Binned & $0.53 \pm 0.29$ & $6.9 \pm 2.5$ & $0.2 \pm 1.3$ \\
Integrated & $1.48 \pm 0.94$ & -- & --\\
\midrule
& \multicolumn{3}{c}{$\mathbf{E}$} \\
Binned & $-4.1 \pm 4.2$ & -- & --\\
\midrule
& \multicolumn{3}{c}{$\mathbf{T}$ + $\mathbf{E}$} \\
Binned & $0.64 \pm 0.23$ & -- & -- \\
\bottomrule\bottomrule
\end{tabular}
\tablefoot{See Table~\ref{tab:png} for more details. Note that the amplitudes of the unclustered point sources $b_\mathrm{PS}$ and CIB $b_\mathrm{CIB}$ bispectra are estimated jointly.}
\end{table}

In Table~\ref{tab:errors}, we illustrate the improvement in constraints achieved with the PR4 dataset compared to the PR3 results (using \texttt{SEVEM} for both). First, we show the expected refinement computed with a Fisher approach. This improvement ranges from a few up to a dozen percent, depending on the shape and if temperature and/or polarization is considered. In the most constraining case, which is the binned bispectrum T+E analysis, smaller error bars (by $7$, $4$ and $8\%$ for the local, equilateral and orthogonal shapes, respectively, and $11\%$ for lensing) are predicted. Second, we do the same comparison between error bars determined in the actual analyses using the binned bispectrum and integrated bispectrum estimators on the \Planck\ PR4 and PR3 datasets. In most cases, they are very close to the Fisher forecast. For the binned bispectrum T+E analysis, the improvement is however slightly lower than expected for the local shape, and higher for the orthogonal and lensing shapes, with a difference of $4\%$ each time. This small difference can be explained by the fact that the error bars themselves are determined with a relative error of $4.1\%$ and $2.8\%$ for PR3 and PR4,\footnote{The relative error on the standard deviation is given by $1/\sqrt{2(n-1)}$ where $n$ is the number of simulations used ($300$ and $600$ for PR3 and PR4, respectively).} respectively. Another important result to mention is that all estimator error bars are compatible with their Fisher predictions. Only for the lensing shape do we obtain results that are slightly suboptimal, with a smaller difference in the case of PR4 ($15\%$ in PR4 vs $21\%$ in PR3 for the binned bispectrum T+E).

\begin{table}[htbp!]   
\centering
\caption{Error bar comparison between \Planck\ PR3 and PR4}
\label{tab:errors}
\renewcommand{\arraystretch}{1.2} 
\setlength{\tabcolsep}{8pt} 
\begin{tabular}{l|cccc}
\toprule\toprule
 & \multicolumn{3}{c}{Binned} & \multicolumn{1}{c}{Integrated} \\
\cmidrule(lr){2-4} \cmidrule(lr){5-5}
 & $T$ & $E$ & $T$ + $E$ & $T$ \\
 \midrule
& \multicolumn{4}{c}{$\mathbf{Fisher}$} \\
Local & $0.96$ & $0.83$ & $0.93$ & $0.98$  \\
Equilateral & $0.98$ & $0.94$ & $0.96$ & --  \\
Orthogonal & $0.92$ & $0.88$ & $0.92$ & --  \\
Lensing & $0.88$ & $0.76$ & $0.89$ & $0.95$  \\
Point sources & $0.88$ & -- & -- & -- \\
CIB & $0.91$ & -- & -- & --  \\
\midrule
& \multicolumn{4}{c}{$\mathbf{Observed}$} \\
Local & $0.96$ & $0.81$ & $0.97$ & $0.94$  \\
Equilateral & $1.03$ & $0.95$ & $0.96$ & --  \\
Orthogonal & $0.90$ & $0.88$ & $0.88$ & --  \\
Lensing & $0.88$ & $0.69$ & $0.85$ & $0.89$  \\
Point sources & $0.86$ & -- & -- & -- \\
CIB & $0.90$ & -- & -- & --  \\
\bottomrule\bottomrule
\end{tabular}
\tablefoot{Ratio of the bispectrum amplitude parameter error bars (PR4 divided by PR3, using \texttt{SEVEM} for both) for the different shapes considered in Tables~\ref{tab:png} and~\ref{tab:isw-lensing}.}
\end{table}

These different results confirm the tightening of the constraints on the NG amplitude parameters obtained from a bispectrum analysis of the \Planck\ PR4 data. The few percent improvement for the primordial local, orthogonal, and equilateral constraints makes them the most stringent constraints on these parameters to date, without changing the main conclusions of the previous \Planck\ PNG analysis based on the PR3 data \citep{Planck:2019kim}.

\section{Validation tests}
\label{sec:tests}

To confirm the robustness of our data analysis results presented in Sect.~\ref{sec:results}, we apply the same pipelines to the \Planck\ PR4 simulations. In Table~\ref{tab:simulations}, we show the amplitude parameters averaged over the $600$ available simulations and their corresponding standard errors, for the primordial local, equilateral and orthogonal templates, and the lensing shape.

We verify that the PNG $\fnl$'s are consistent with the expected value of zero, as no PNG is included in the simulations. While there are small deviations from zero (of order $1$ to $1.5\sigma$) in the T or E-only analyses for the local and equilateral shapes, they disappear in the binned T+E case, which provides the most stringent constraints. The situation is similar for the orthogonal shape, which however has a slightly stronger deviation ($2.5\sigma$) in the binned T-only analysis. Even if this would turn out not to be a statistical fluctuation, this deviation is still one order of magnitude smaller than the error bar on one map, used for the constraints in Table~\ref{tab:png}, and thus cannot bias our results in any significant manner.

Concerning the lensing bispectrum, we find a significant deviation from the expected value of $1$, with both estimators and in all cases (although the significance is lower for the binned E-only and integrated bispectrum analyses, due to much larger error bars). In the remainder of the section, we check several possible origins of this issue and verify that it has no impact on our observational constraints derived from the \Planck\ PR4 data. However, we want to stress that even if this deviation is significant statistically due to the averages computed from many simulations, it still represents only a small fraction of the $1\sigma$ observational error bars and is therefore negligible with respect to the statistical fluctuations expected in the data maps.

\begin{table*}[htbp!]
\centering
\caption{Bispectrum amplitude parameters in the $600$ \Planck\ PR4 simulations}
\label{tab:simulations}
\renewcommand{\arraystretch}{1.2} 
\setlength{\tabcolsep}{8pt} 
\begin{tabular}{l|cccc}
\toprule\toprule
 & \multicolumn{3}{c}{Binned} & \multicolumn{1}{c}{Integrated} \\
\cmidrule(lr){2-4} \cmidrule(lr){5-5}
 & $T$ & $E$ & $T$ + $E$ & $T$ \\
 \midrule
Local & $-0.35 \pm 0.23$ & $0.19 \pm 0.96$ & $0.02 \pm 0.20$ & $-0.21 \pm 0.30$  \\
Equilateral & $3.2 \pm 2.9$ & $-9.0 \pm 6.1$ & $1.1 \pm 1.9$ & --  \\
Orthogonal & $3.4 \pm 1.4$ & $-4.8 \pm 3.1$ & $1.0 \pm 0.9$ & --  \\
Lensing & $0.925 \pm 0.012$ & $0.71 \pm 0.17$ & $0.939 \pm 0.009$ & $0.90 \pm 0.04$ \\
\bottomrule\bottomrule
\end{tabular}
\tablefoot{The amplitude parameters are determined with the same estimators as the data analyses presented in Sect.~\ref{sec:results} from the $600$ \texttt{SEVEM} PR4 simulations. We show the average values, and the corresponding standard errors. The lensing bias is subtracted from the three PNG shapes (local, equilateral and orthogonal).}
\end{table*}

We have performed a series of tests to confirm that this issue is already present at the level of the original CMB simulations and thus is neither an artifact of the \Planck\ PR4 NPIPE pipeline nor of the \texttt{SEVEM} component separation technique. First, we do a direct comparison of the \Planck\ PR3 and PR4 simulation results. For both releases, lensed CMB realizations have been produced and processed through different pipelines to reproduce as accurately as possible the different instrumental effects (see \cite{Planck:2018bsf, Planck:2018lkk, Planck:2018yye} and \cite{Planck:2020olo} for details on PR3 and PR4, respectively). There is a common set of $100$ CMB seeds fully processed for both releases (the ones numbered from $200$ to $299$). 

In Table~\ref{tab:simulations-100}, we show the results obtained from the analysis of these $100$ simulations for both PR3 and PR4 using the binned bispectrum estimator. Concerning the local, equilateral and orthogonal shapes, the measured $\fnl$ values are compatible with the expected absence of PNG in these simulations. The lensing bispectrum is measured with similar amplitudes for the T and T+E amplitudes in both sets close to a value of $0.9$, a few $\sigma$ below the expected value of $1$ (the E-polarization error bars are one order of magnitude larger and thus not sufficient to detect any deviation of the same order). This means that the new processing of instrumental effects used in the PR4 simulations does not bias the lensing results, at least with respect to the PR3 results. Note that in this specific set of $100$ simulations, most error bars are of the same order between the two releases. This is because $100$ simulations are not sufficient to obtain fully converged numerical results that can show the small improvements obtained with PR4 that we highlighted in Sect.~\ref{sec:results}.

\begin{table*}[htbp!]
\centering
\caption{Bispectrum amplitude parameters in the $100$ common \Planck\ PR3 and PR4 simulations (\texttt{SEVEM})}
\label{tab:simulations-100}
\renewcommand{\arraystretch}{1.2}
\setlength{\tabcolsep}{8pt} 
\begin{tabular}{l|cccccc}
\toprule\toprule
 & \multicolumn{3}{c}{PR3} & \multicolumn{3}{c}{PR4} \\
\cmidrule(lr){2-4} \cmidrule(lr){5-7}
 & $T$ & $E$ & $T$ + $E$ & $T$ & $E$ & $T$ + $E$\\
 \midrule
Local & $-0.54 \pm 0.60$ & $-3.6 \pm 3.1$ & $-1.00 \pm 0.56$ & $0.19 \pm 0.56$ & $-4.5 \pm 2.4$ & $0.39 \pm 0.52$   \\
Equilateral & $2.7 \pm 7.1$ & $-19 \pm 16$ & $-0.9 \pm 4.5$ & $ 0.7 \pm 7.2$ & $-19 \pm 15$ & $1.7 \pm 4.6$   \\
Orthogonal & $-2.8 \pm 3.3$ & $9.9 \pm 9.2$ & $1.2 \pm 2.5$ & $0.4 \pm 3.1$ & $1.0 \pm 7.1$ & $0.5 \pm 2.2$   \\
Lensing & $0.898 \pm 0.030$ & $ 1.37 \pm 0.65$ & $0.907 \pm 0.025$ & $0.893 \pm 0.032$ & $0.66 \pm 0.41$ & $0.927 \pm 0.025$ \\
\bottomrule\bottomrule
\end{tabular}
\tablefoot{The amplitude parameters are determined with the binned bispectrum estimator from the $100$ \texttt{SEVEM} PR3 and PR4 simulations that share the same CMB seeds. We show the average values, and the corresponding standard errors. The lensing bias is subtracted from the three PNG shapes
(local, equilateral and orthogonal).}
\end{table*}

We also evaluate the impact of the cleaning technique \texttt{SEVEM} on the PNG results. We have to focus on the PR3 simulated maps, as more component separation methods were applied for that release. In \cite{Planck:2019kim}, the main reported results were based on the method \texttt{SMICA} \citep{Cardoso:2008qt}. In Table~\ref{tab:simulations-smica} we verify that on the $300$ available PR3 simulations, we obtain very similar results. By comparison of Tables~\ref{tab:simulations-100} and \ref{tab:simulations-smica}, we directly see that the results are extremely close to each other, at a level much better than the expected error bars. As the two cleaning procedures are based on very different methods, no correlation of possible biases is expected, and thus this confirms that neither the PNG nor the lensing bispectrum amplitudes are biased by \texttt{SEVEM}.

\begin{table*}[htbp!]
\centering
\caption{Bispectrum amplitude parameters in the $300$ \Planck\ PR3 simulations with two different component separation techniques}
\label{tab:simulations-smica}
\renewcommand{\arraystretch}{1.2} 
\setlength{\tabcolsep}{8pt} 
\begin{tabular}{l|cccccc}
\toprule\toprule
  & \multicolumn{3}{c}{\texttt{SEVEM}} & \multicolumn{3}{c}{\texttt{SMICA}} \\
\cmidrule(lr){2-4} \cmidrule(lr){5-7}
 & $T$ & $E$ & $T$ + $E$ & $T$ & $E$ & $T$ + $E$ \\
 \midrule
Local & $0.09 \pm 0.33$ & $-0.02 \pm 1.7$ & $-0.70 \pm 0.30$ & $0.06 \pm 0.32$ & $0.01 \pm 1.5$ & $-0.72 \pm 0.29$ \\
Equilateral & $ 6.1 \pm 4.0$ & $-9.9 \pm 9.1$ & $ 0.4 \pm 2.8$ & $5.6 \pm 4.0$ & $-5.7 \pm 8.4$ & $0.3 \pm 2.8$ \\
Orthogonal & $-0.5 \pm 2.3$ & $ 9.3 \pm 5.1$ & $2.6 \pm 1.4$ & $-0.1 \pm 2.3$ & $7.1 \pm 4.8$ & $2.9 \pm 1.4$ \\
Lensing & $0.898 \pm 0.030$ & $ 1.37 \pm 0.65$ & $0.960 \pm 0.015$ & $0.977 \pm 0.019$ & $0.85 \pm 0.30$ & $0.962 \pm 0.015$ \\
\bottomrule\bottomrule
\end{tabular}
\tablefoot{The amplitude parameters are determined with the binned bispectrum estimator from the $300$ \texttt{SMICA} and \texttt{SEVEM} PR3 simulations. We show the average values, and the corresponding standard errors. The lensing bias is subtracted from the three PNG shapes (local, equilateral and orthogonal).}
\end{table*}

A further investigation of this small discrepancy between the measured lensing amplitude in the simulations and the expected value of $1$ would require to check in detail steps like the lensing implementation in the CMB simulations based on \texttt{LensPix} \citep{Lewis:2005tp} and the accuracy of the theoretical computation of the lensing bispectrum. This is however beyond the scope of this work and thus we focus here on confirming that this discrepancy does not have an impact on the reported observational constraints.

The binned and integrated bispectrum estimators explicitly reconstruct the bispectrum in addition to measuring amplitude parameters. This allows us to use the averaged bispectrum from the simulations as a theoretical template, to estimate how much it can bias the results from the observations. In Table~\ref{tab:png2}, we show the $\fnl$ parameters of the local, equilateral, and orthogonal shapes measured in the PR4 data, after subtracting the bias of the averaged simulation bispectrum. As can be verified by comparison to Table~\ref{tab:png}, these results are extremely close to the values obtained by removing the standard lensing theoretical bias, with deviations one order of magnitude smaller than the $1\sigma$ error bars. This confirms that even if there is a small discrepancy between the simulations and the expected theoretical predictions, it is much too small to affect the reported observational constraints, at least as far as the central values are concerned.

\begin{table}[htbp!]   
\centering
\caption{Constraints on PNG from \Planck\ PR4 with different bias subtraction}
\label{tab:png2}
\renewcommand{\arraystretch}{1.2} 
\setlength{\tabcolsep}{8pt}
\begin{tabular}{l|ccc}
\toprule\toprule
\cmidrule(lr){2-4} 
Estimator & Local & Equilateral & Orthogonal \\
\midrule
& \multicolumn{3}{c}{$\mathbf{T}$} \\
Binned & $0.9 \pm 5.5$ & $20 \pm 72$ & $12 \pm 35$  \\
Integrated &  $7.4 \pm 7.3$ & --  & --  \\
\midrule
& \multicolumn{3}{c}{$\mathbf{E}$} \\
Binned & $16 \pm 24$ & $23 \pm 150$ & $-30 \pm 77$  \\
\midrule
& \multicolumn{3}{c}{$\mathbf{T}$ + $\mathbf{E}$} \\
Binned & $-0.6 \pm 5.0$ & $5 \pm 46$ & $-9 \pm 21$ \\
\bottomrule\bottomrule
\end{tabular}
\tablefoot{Similar to the right-hand side of Table~\ref{tab:png}, but with a different lensing bias subtraction. Here the bias due to the averaged bispectrum of the $600$ PR4 CMB \texttt{SEVEM} simulations has been subtracted from each reported PNG amplitude.}
\end{table}

\begin{figure*}
    \centering
    \includegraphics[width=0.99\linewidth]{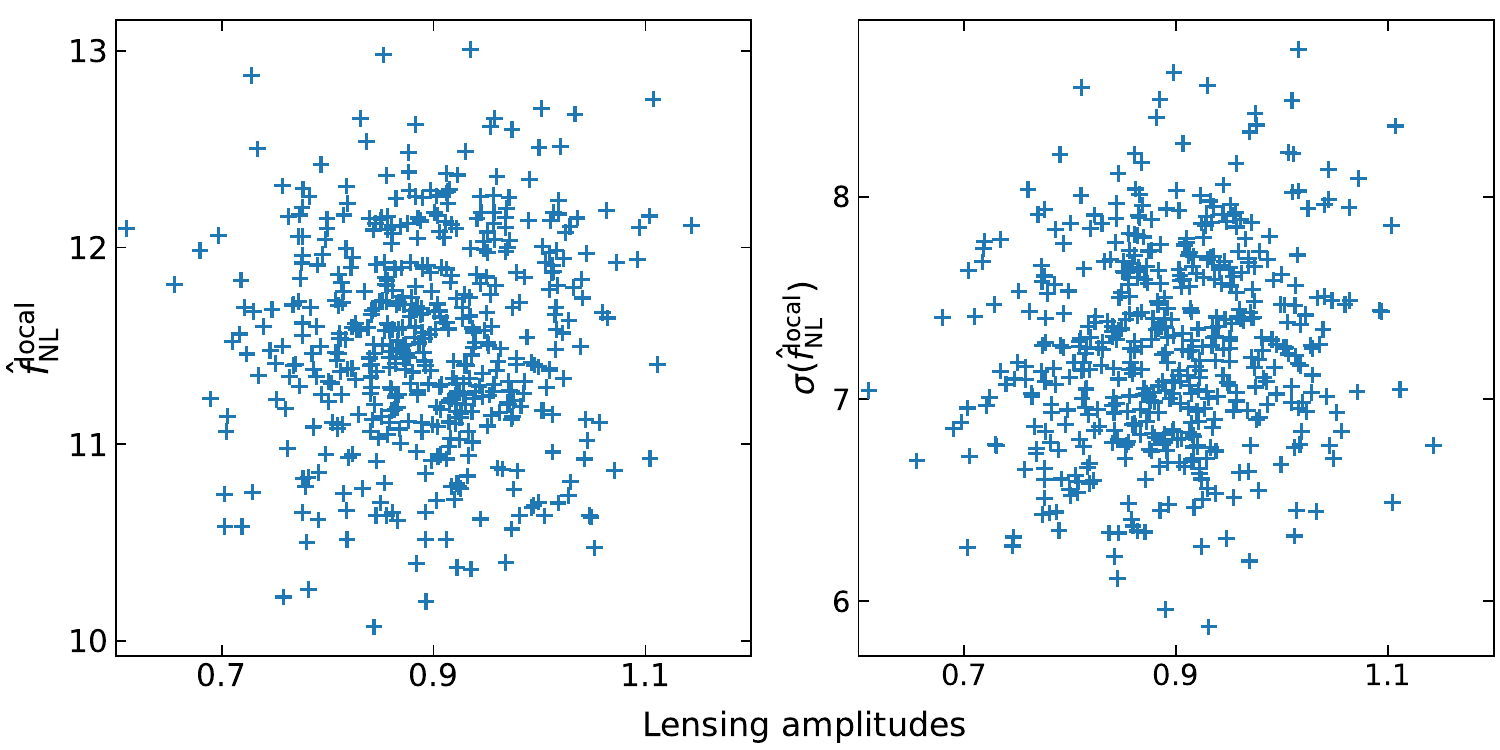}
    \caption{Estimated $\hat{f}_\mathrm{NL}^{local}$ (left) and its error bar $\sigma(\hat{f}_\mathrm{NL}^{local})$ (right) using different subsets of the PR4 simulations for linear correction and error bars, as a function of the lensing amplitude in the corresponding subset, determined using the integrated bispectrum estimator. Each subset consists of $100$ maps drawn randomly from the $600$ \texttt{SEVEM} PR4 simulations. Note that the expected lensing bias of $4.3$ on the estimated $\hat{f}_\mathrm{NL}^{local}$ has not been subtracted.}
    \label{fig:isw-local}  
\end{figure*}

We now check the impact on the error bars. We focus on the local PNG shape, which is by far the most correlated with the lensing bispectrum as they both peak in the squeezed limit. We use the integrated bispectrum estimator with subsets of $100$ simulations taken among the $600$ available ones to compute both the linear correction term and the error bars (for temperature alone). We randomly draw many of such subsets and measure the average lensing amplitude of the simulations, as well as the central value and error bar of $\fnl^\mathrm{local}$. For all these sets, we show the corresponding values in Fig.~\ref{fig:isw-local}. Although there is a variation of both $\fnl^\mathrm{local}$ and its error bar depending on the specific subset, the results confirm that it is uncorrelated to the lensing amplitude of the specific subset and that it is actually compatible with the differences expected from a random Gaussian field. 

\section{Conclusions}
\label{sec:conclusions}
In this work we performed the first analysis of the standard non-Gaussian bispectrum templates
on the \Planck\ PR4 CMB data \citep{Planck:2020olo}, both temperature and E-polarization. 
These are the local,
equilateral and orthogonal primordial templates, and the secondary bispectrum templates
for lensing, unclustered extra-galactic point sources and the CIB. The results of the
full T+E analysis are $\fnl^\mathrm{local} = -0.1 \pm 5.0$, $\fnl^{equil} = 6 \pm 46$ and
$\fnl^\mathrm{ortho} = -8 \pm 21$. While the improvements of the error bars are only 
a few percent with respect to the PR3 results (up to $12\%$ for orthogonal), these now
stand as the best constraints on primordial non-Gaussianity from \Planck. As the error
bars were determined from 600 simulations instead of 300 for PR3, they should also be 
even more accurate. Results are fully consistent with the PR3 results and no primordial 
non-Gaussianity is detected.

The analysis in this paper is necessarily more limited than the one in the full \Planck\ 
collaboration paper on primordial non-Gaussianity for PR3 \citep{Planck:2019kim}. In the 
first place this is due
to the available maps: only the \texttt{SEVEM} component separation technique was applied
to the PR4 maps to produce everything needed for a non-Gaussianity analysis. For PR3
four different component separation techniques were applied, including \texttt{SMICA} that
was chosen as the default method in all previous releases. In the second place this is
due to using only one effectively optimal bispectrum estimator, the binned bispectrum 
estimator, with support from the non-optimal but very fast integrated bispectrum estimator 
for cross-checking certain results and running additional tests. For PR3 three different
effectively optimal estimators (with four independent pipelines) were used, including the 
KSW estimator
that was chosen as the default for the three primordial shapes in the previous releases.
In fact we originally planned to have also results with the modal bispectrum estimator
in this paper, but due to time constraints and computational resource limitations this
was not possible in the end. However, it was shown in all previous releases that the
results from the different component separation techniques and the different bispectrum
estimators are fully consistent using many dozens of validation tests, and the binned 
bispectrum pipeline
used here is the same as that used and fully validated in the past releases. In 
addition, as described we performed many validation tests for this paper as well. Hence 
the results should be fully robust.

Regarding the non-primordial templates, as in PR3 we detect lensing and unclustered point 
sources but no CIB (in a joint analysis of the latter two). Also similar to PR3, the
measured lensing signal is a bit low compared to the expected value of 1, although 
compatible with 1 given the size of the error bars. It should be
pointed out that in PR3 the variation between the four component separation techniques was
larger regarding the non-primordial shapes, and \texttt{SEVEM} was the one with the highest
point source signal and the lowest lensing signal. In the PR3 paper it was pointed out
that the low lensing signal might be a consequence of the correlation between
the lensing template and another bispectrum describing the correlation between 
the ISW and tSZ effects. This was made plausible by showing that the lensing signal
was much closer to unity in maps where the tSZ contribution had been removed, although no
final conclusions could be drawn given the size of the error bars and the lack of 
significance. Lacking PR4 maps where the tSZ contribution has been removed, we could not 
redo this check here.

We performed several validation tests, the most interesting of which we report in
this paper. In particular we checked that our pipeline works as expected on the 600
simulations. While we found no PNG in these simulations as expected, we were surprised to
detect a very significant deviation downwards from 1 for the lensing shape. As the
simulations contain no foregrounds, this cannot be due to the tSZ effect. Further
investigation showed that this was not due to NPIPE, the new data-processing pipeline in 
PR4, nor to \texttt{SEVEM}, but was already present in the original CMB seed simulations 
for PR3 (that were reused in PR4). Due to a fortuitous choice of simulations for the PR3 
analysis, this effect was not as notable at the time. However, we 
show that this mismatch of the lensing bispectrum in the simulations has no impact on our
reported results for the CMB data map, neither regarding the central values nor the 
error bars.

PNG remains one of the most important sources of information about inflation and the
early universe in general. Hence it is important to improve our constraints and their robustness,
as we did in this paper, even if the improvement is modest. While undetected so far, 
that non-detection of PNG has already ruled out many models of inflation and alternatives. 
New observational data, from CMB or large-scale structure, whether they lead to a detection or a further 
tightening of the constraints, can only improve our knowledge of the early universe.
PNG has the additional advantage that, unlike the tensor-to-scalar ratio, the prediction
from single-field slow-roll inflation provides a clear lower limit to $\fnl$.
And while waiting for new experiments to come online and provide data, the wealth of
information represented by the \Planck\ data is here and, who knows, might still contain 
some surprises regarding as yet untested bispectrum templates\ldots

\begin{acknowledgements}
We thank Michele Liguori for useful inputs and discussions.
GJ and NA acknowledge support from the ANR LOCALIZATION project, grant ANR-21-CE31-0019 / 490702358 from the French Agence Nationale de la Recherche / DFG.
This research made use of observations obtained with \Planck\ (\url{http://www.esa.int/Planck}), an ESA science mission with instruments and contributions directly funded by ESA Member States, NASA, and Canada.
The authors acknowledge the use of the {\em healpy} and {\em HEALPix} packages \citep{Zonca2019, 2005ApJ...622..759G}.
We gratefully acknowledge the IN2P3 Computer Centre (\url{https://cc.in2p3.fr}) for providing the computing resources and services needed for the analysis with the binned bispectrum estimator.
\end{acknowledgements}


\bibliographystyle{aa}
\bibliography{biblio}


\end{document}

%% file: preamble.tex
\newcommand{\ib}{\mathrm{I}\!B}
\newcommand{\ic}{\mathrm{IC}}
\newcommand{\fnl}{f_\mathrm{NL}}

\newcommand{\fsky}{f_\mathrm{sky}}
\newcommand{\lmax}{\ell_\mathrm{max}}

\newcommand{\Planck}{\textit{Planck}}